\documentclass[
twocolumn,
 amsmath,amssymb,
 aps,
pra,
 showpacs
]{revtex4-1}
\usepackage{placeins}
\usepackage{amsmath}
\usepackage{mathrsfs}
\usepackage{graphicx}
\usepackage{dcolumn}
\usepackage{bm}
\usepackage{ulem}
 \usepackage{relsize}
\usepackage[
margin=.7in,
]{geometry}

\begin{document} 
\title{Identifying spatially asymmetric high-order harmonic emission during the falling edge of an intense laser pulse
}
\author{M. Vafaee$^{1}$}
\author{H. Ahmadi$^{2}$ }
\author{A. Maghari$^{2}$}

\affiliation{
$^{1}$Department of Chemistry, Tarbiat Modares University, P. O. Box 14115-175, Tehran, Iran
\\$^{2}$Department of Physical Chemistry, School of Chemistry, College of Science, University of Tehran, Tehran, Iran
 }

\begin{abstract}
Two different induced effects  of a laser falling edge on high-order harmonic generation are resolved by solving numerically full-dimensional electronic time-dependent Schr\"{o}dinger equation beyond the Born-Oppenheimer approximation. The harmonic spectrum of H$_2^+$ and T$_2^+$ isotopes are compared to see the effects of a 4-cycle falling edge of a 800 nm, 15-cycle trapezoidal laser pulse of $I=$3 $\times 10^{14}$ Wcm$^{-2}$ intensity on harmonic emission spectrum. The harmonic emission at the laser falling part is negligible for H$_2^+$ due to ionization suppression, but considerable for T$_2^+$. The falling edge of  the laser pulse induces two effects on the HHG in T$_2^+$. The first well-known effect is non-adiabatic frequency redshift of generated odd-order harmonics. The second unknown one is spatially asymmetric harmonic emission which appears as even harmonic orders. In order to clarify this new effect, spatial distribution of HHG and resolving HHG into different components are demonstrated. The asymmetric emission would appear for both atoms and molecules as long as harmonic emission of either rising or falling edge of an intense trapezoidal or non-trapezoidal laser pulse is dominant.  
\end{abstract}


\maketitle
\section{Introduction}
An electron in the ground state of  hydrogen atom in Bohr's model has an orbital period of 150 attosecond (1 as=$10^{-18}$ s). In order to temporally resolve electronic dynamics in real time, it is necessary to follow its dynamics within attosecond time scales. Nowadays, this  attosecond temporal resolution is accessible with attosecond and even femtosecond laser pulses [1]. First attosecond laser pulse was measured in 2001 based on high-oder harmonic generation (HHG)[2,3].
 The underlying mechanism of HHG is often described by a three-step semi-classical model [4]. In the first step, an electron releases into continuum via tunneling. Then, the electron moves away from the ion, but it is driven back when the field laser sign reverses. Eventually, energetic photons are emitted via the recombination of the released electron with its parent ion. The three-step model predicts maximum recollision energy of $3.17U_p$, where $U_p=I/{4\omega^2}$, is the pondermotive energy in which $I$ and $\omega$ are laser intensity and angular frequency, respectively. The improved version of the three-step model has been also introduced by Lewenstein \textit{et al}.  by incorporating quantum mechanical corrections [5].

The characteristics of each emitted photon in the HHG process are determined by both the driving laser field and the generating medium. 
For example, for Gaussian-like laser pulses having  rising and falling parts, the effective amplitude of each cycle experienced by atoms and molecules changes non-adiabatically from a laser cycle to  another. Thus, the energy  acquired in each cycle by the freed electron is different. This non-adiabatic response of a medium to a rapidly changing laser field leads to a frequency red-shift (blue-shift) of harmonics at the falling (rising) part of the laser pulse [6-9].

In practice, it is difficult to observe purely the non-adiabatic redshift on the laser falling edge in atoms as the blue-shift usually occurs at the laser rising edge. When the laser intensity is above the saturation intensity threshold, $I_s$, most harmonic emission occurs at the laser rising part which means that only blue-shifted harmonics are expected to appear. In other words, HHG is suppressed considerably at the laser falling edge due to a high population depletion at the rising part.  If the laser intensity is below $I_s$, the non-adiabatic blueshift and redshift are comparable. Thereby, the spectrum of each harmonic broadens and no net shift is observed. The non-adiabatic blue-shift of the harmonics has been reported experimentally [10,11], but there is no experimental report on the non-adiabatic redshift for both atoms and molecules. 

In molecules because of  more degrees of freedom than atoms, it is possible to observe both the non-adiabatic redshift and blueshift under appropriate conditions.
It is shown that the nuclear motion can serve as a controllable tool to trap the non-adiabatic redshift in molecules for the Gaussian [12] and trapezoidal [13] laser pulses. The ionization and recombination processes  can be controlled to occur dominantly on the trailing edge of the laser pulse by choosing  appropriate laser parameters and molecule. In fact, the time-dependent ionization of molecules permits the observation of the non-adiabatic redshift. References [12,13] are focused on the non-adiabatic redshift induced by the falling edge of the laser pulse. 

In this work, we similarly control the harmonic emission at the falling laser part by choosing a proper pulse shape and molecules. For this purpose, isotopes H$_2^+$ and T$_2^+$ under a 15-cycle trapezoidal laser are investigated. We show that harmonic emission at the 4-cycle laser falling part is negligible for H$_2^+$ due to the ionization suppression, but it is considerable for T$_2^+$. Thus, we attribute the difference of the HHG spectrum between these two isotopes to the induced effects of the falling edge.  We identify spatially asymmetric emission  as a distinct induced effect of the falling edge. The spatially asymmetric emission appears as even harmonic orders as a result of the spatial symmetry breaking. 
 
 In this report, our focus  is mainly to introduce  and identify spatially  asymmetric emission. To our knowledge, this kind of spatially asymmetric emission has not been addressed and clarified  in the  literature. In order to understand deeply the underlying physics behind the asymmetric emission, HHG is spatially resolved and decomposed into different localized signals.

In this work, three-dimensional  electronic time-dependent Schr\"{o}dinger equation (TDSE) beyond the Born-Oppenheimer approximation (NBO) is numerically solved  for H$_2^+$ and T$_2^+$. Calculations have been done with 15-cycle trapezoidal laser pulses at 800 nm wavelength and $I=$3 $\times 10^{14}$ Wcm$^{-2}$ intensity.  
 We assume that molecular ions are aligned with their internuclear-distance axis parallel to the laser polarization direction. The molecular alignment is frequently implied experimentally nowadays [14-16]. We use atomic units throughout the article unless stated otherwise.

\section{Computational Methods}
The time-dependent Schr\"{o}dinger equation for homonuclear hydrogen-like molecular ions with the electron's  coordinates  $z$ and $\rho$ can be written (after a separation of the center-of-mass motion and ignoring molecular rotations) as [17-18]

\begin{eqnarray}\label{eq:1}
  i \frac{\partial \psi(z, \rho, R;t)}{\partial t}={\widehat H}(z, \rho, R;t)\psi(z, \rho, R;t),
\end{eqnarray}
 where $R$ denotes internuclear distance (parallel to both electronic coordinate $z$ and the laser polarization direction) and \^{H} is the total electronic and nuclear Hamiltonian which is given by
\begin{eqnarray}\label{eq:2}
 \widehat{H}(z, \rho, R;t)=&\mathlarger{-\frac{2m_N+m_e}{4m_Nm_e}[\frac{\partial^2}{\partial z^2}}+\frac{\partial^2}{\partial \rho^2}+\frac{1}{\rho}\frac{\partial}{\partial \rho}]
\nonumber \\
& \mathlarger{-\frac{1}{m_N}\frac{\partial^2}{\partial R^2}+V_C(z, \rho, R;t)},
\end{eqnarray}
with
\begin{eqnarray}\label{eq:3}
 V_C&(z, \rho, R;t)=\mathlarger{-\frac{1}{\sqrt{(z+\frac{R}{2})^2+ \rho^2}}-\frac{1}{\sqrt{(z-\frac{R}{2})^2+ \rho^2}}}
\nonumber \\
 &\mathlarger{+\frac{1}{R}+(\frac{2m_N+2m_e}{2m_N+me})zE_0f(t)cos(\omega t)}.
\end{eqnarray}
In the above equations, $E_0$ is the laser peak amplitude, $m_e$ and $m_N$ are  masses of the electron and single nucleus, respectively, $\omega$ angular frequency,  and \textit{f}(t) is the laser pulse envelope. The envelope  rises linearly during the first four cycles, then remains constant for seven cycles and decreases during the last four cycles.

\begin{figure*}[ht]
\begin{center}
\begin{tabular}{c}
\centering
\resizebox{165mm}{60mm}{\includegraphics{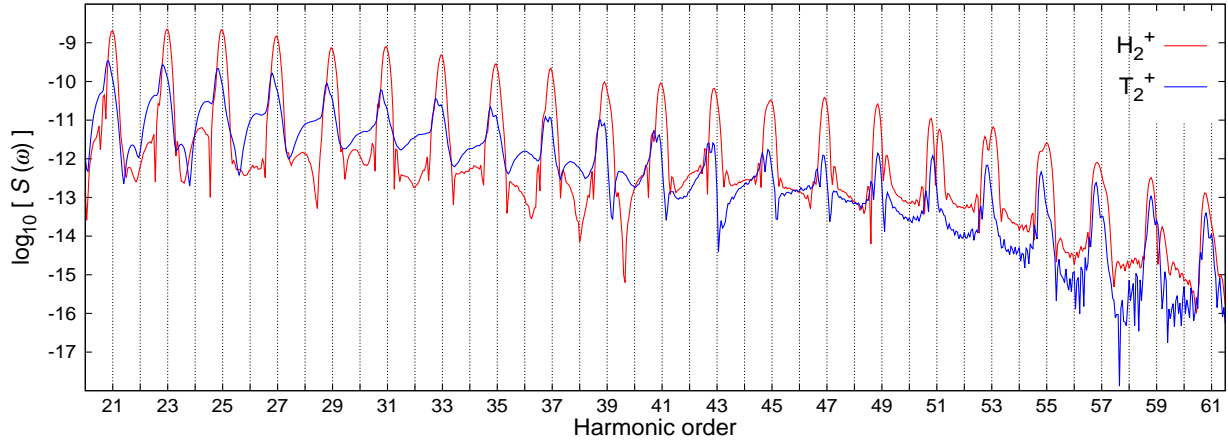}}
\end{tabular}
\caption{
\label{HHG} 
(Color online) High-order harmonic spectra produced by H$_2^+$ (red) and T$_2^+$ (blue) under 15-cycle trapezoidal laser pulse of 800 nm wavelength and $I=$3 $\times 10^{14}$ Wcm$^{-2}$ intensity.		}
\end{center}
\end{figure*}
\begin{figure}[ht]
\centering
\resizebox{85mm}{68mm}{\includegraphics{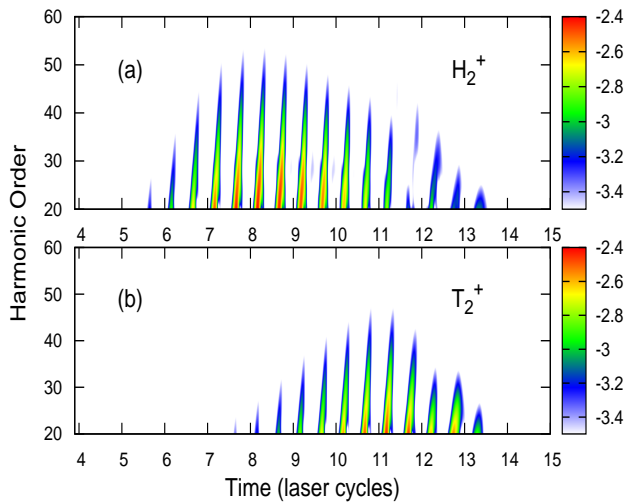}}
\caption{
\label{Morlet-wavelet}
(Color online) The Morlet-wavelet time profiles for H$_2^+$ (a) and  T$_2^+$ (b). The HHG intensities are depicted in color logarithmic scales on the right side of panels. Laser parameters are the same as Fig. 1.
}
\end{figure}
\begin{figure*}[ht]
\begin{center}
\begin{tabular}{l}
\resizebox{145mm}{55mm}{\includegraphics{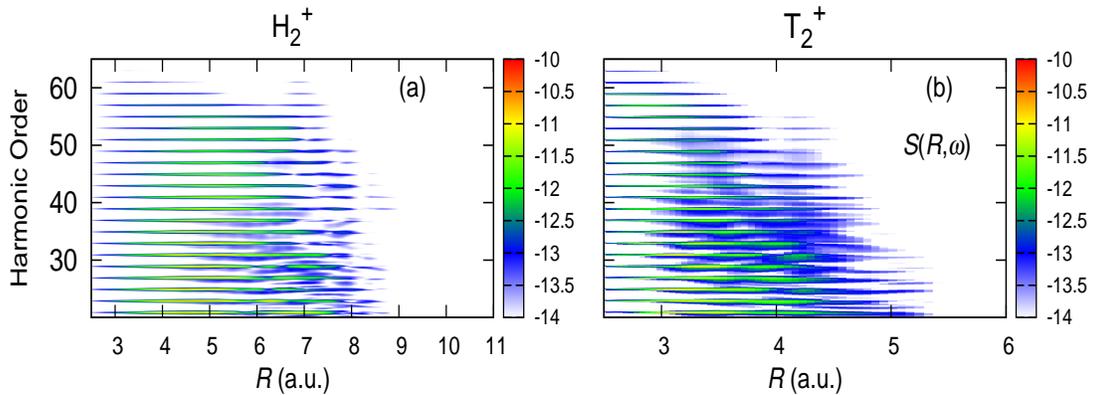}}
\end{tabular}
\caption{
\label{<S(R,w)>}
(Color online) Spatial distribution of the HHG spectra in terms of internuclear distance $R$ and harmonic order for H$_2^+$ (left panel) and T$_2^+$ (right panel).  The HHG intensities are depicted in color logarithmic scales on the right side of the panels.	Laser parameters are the same as Fig. 1.	
		}
\end{center}
\end{figure*}

\begin{figure*}[ht]
\centering
\resizebox{165mm}{100mm}{\includegraphics{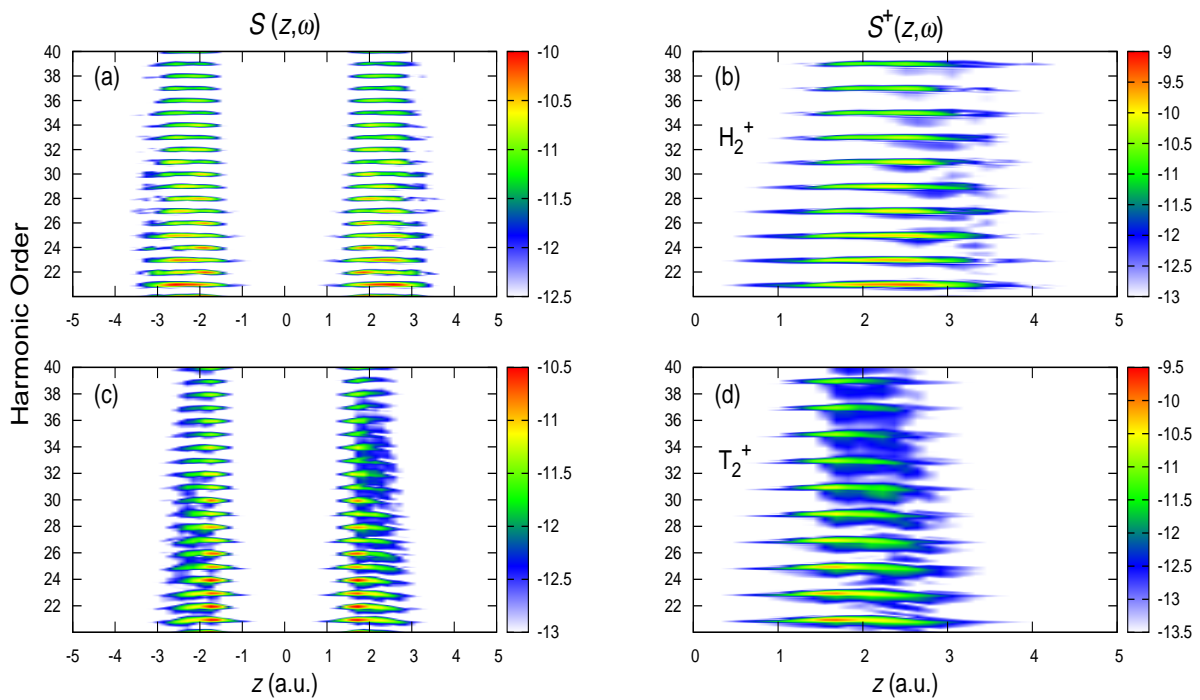}}
\caption{
\label{S(z,w)}
(Color online) Spatial distributions of the HHG spectra in terms of $z$ coordinate and harmonic order, $S(z,\omega)$ (left panels) and $S^+(z,\omega)$ (right panels), for H$_2^+$ (top row) and T$_2^+$ (bottom row).  The HHG intensities are depicted in color logarithmic scales on the right side of the panels.	Laser parameters are the same as Fig. 1. 
}
\end{figure*}

\begin{figure*}[ht]
\centering
\resizebox{145mm}{110mm}{\includegraphics{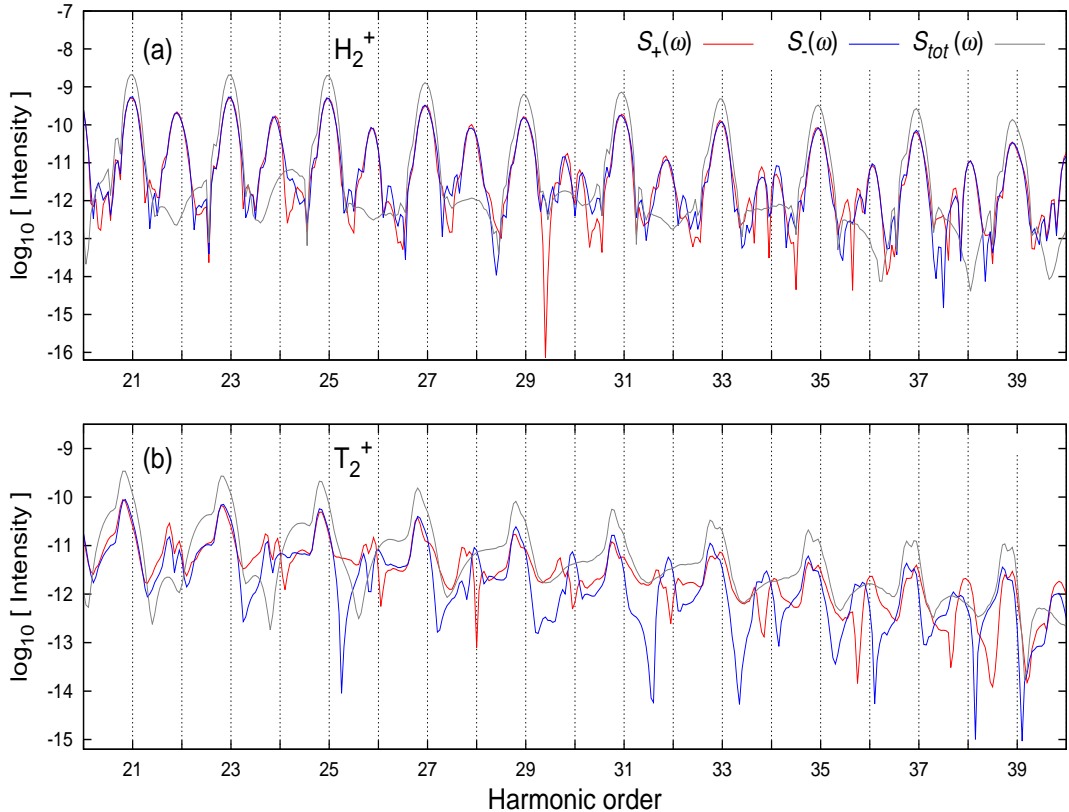}}
\caption{
\label{S_+}
(Color online) First row: $S_{tot}(\omega)$, $S_+(\omega)$ and $S_-(\omega)$ for H$_2^+$ (left panels) and  T$_2^+$ (right panels). Second row: corresponding phase of $S_+(\omega)$ and $S_-(\omega)$  of the first row.  Laser parameters are the same as Fig. 1. 
}
\end{figure*}

\begin{figure}[ht]
\centering
\resizebox{85mm}{50mm}{\includegraphics{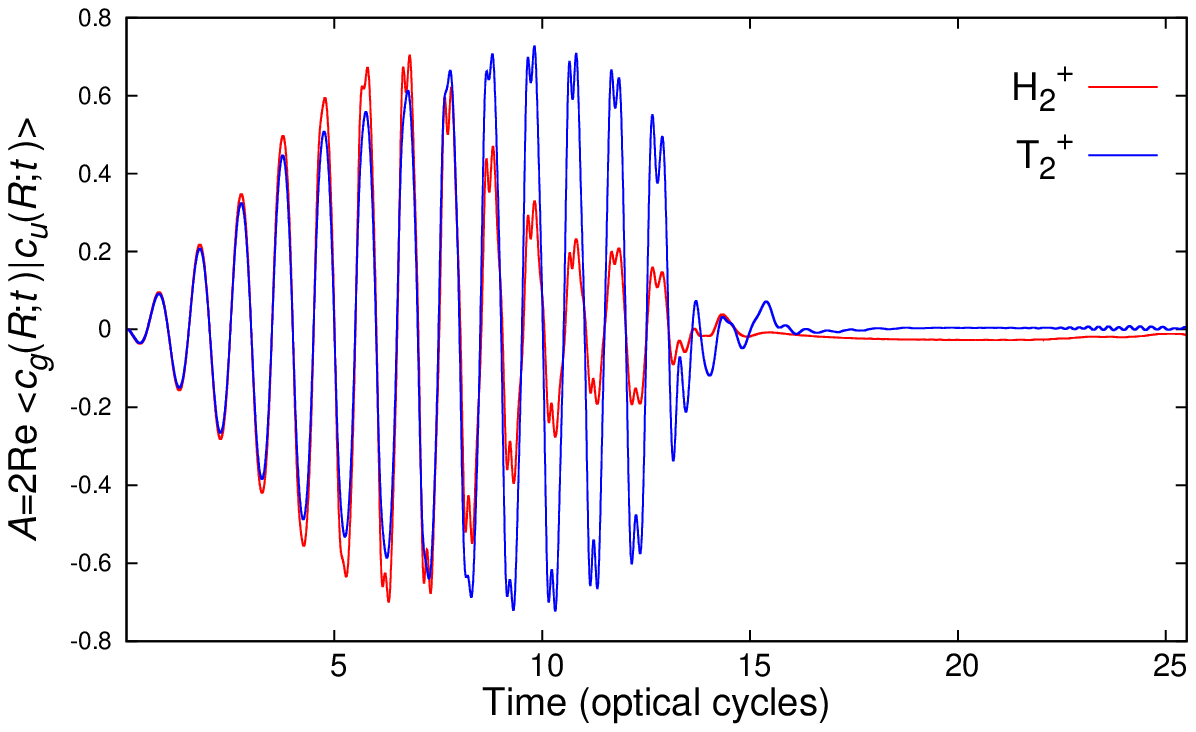}}
\caption{
\label{S_+}
(Color online) Asymmetric parameter $A$ for H$_2^+$ (red curve) and  T$_2^+$ (blue curve).  Laser parameters are the same as Fig. 1. 
}
\end{figure}

The TDSE is numerically integrated with the help of unitary split-operator methods [19-20]. The detail of the numerical procedure is  described in our previous works [21-23]. The finest grid size values in our numerical integration are 0.13, 0.2 and 0.025, respectively for $z$, $\rho$ and $R$ coordinates. The size of the simulation box is chosen as $z_{max}
= 63$, $\rho_{max}= 25$  and $R_{max}= 16$. 
The HHG spectra are calculated as the square of the windowed Fourier transform of dipole acceleration $a_z(t)$ in the electric field direction ($z$) as
\begin{eqnarray}\label{eq:4}
  S(\omega)=&
 \\  
  \mathlarger{\vert}&\int_0^T<\psi(z,\rho, R;t)|a_z(t)\mid \psi(z,\rho, R;t)>_{z,\rho, R}\times \,\nonumber \\ &H(t)\,exp[-i\omega t]\,dt\; \mathlarger{\vert} ^2,\nonumber
\end{eqnarray}
where
\begin{eqnarray}\label{eq:5}
  H(t)= \frac{1}{2}[1-cos(2\pi \frac{t}{T})],
\end{eqnarray}
is the Hanning function and $T$ is the total pulse duration. The Hanning filter reduces the effect of  nondecaying components in the dipole acceleration that last after the laser pulse is switched off [24]. 
The time profile of harmonics is obtained by Morlet-wavelet transform of dipole acceleration $a_z(t)$ via [25-26]
\begin{eqnarray}\label{eq:6}
  &\mathlarger{w(\omega,t)= \sqrt{ \frac{\omega}{\pi^\frac{1}{2}\sigma}}\int_{-\infty}^{+\infty}<\psi(z,\rho, R;t)|a_z(t^\prime)\mid \psi(z,\rho, R;t)>\times}
 \nonumber \\
 &\mathlarger{exp[-i\omega (t^\prime-t)]exp[-\frac{\omega^2 (t^\prime-t)^2}{2\sigma^2}]dt^\prime.}
\end{eqnarray}
We set $\sigma=2\pi$ in this work.

We can easily  express spatial distribution of total HHG spectrum (Eq. (4)) in terms of $R$ as
 
 \begin{eqnarray}\label{eq:7}
  S(R,\omega)=&\\
  \mathlarger{\vert}\int_0^T<\psi(z,\rho, R;t)|a_z(t)&\mid \psi(z,\rho, R;t)>_{z,\rho}\,H(t)\,exp[-i\omega t]\,dt\; \mathlarger{\vert} ^2, \nonumber 
\end{eqnarray} 
 and in terms of $z$ coordinate as
  \begin{eqnarray}\label{eq:8}
  S(z,\omega)=& \\
  \mathlarger{\vert}\int_0^T<\psi(z,\rho, R;t)|a_z(t)&\mid \psi(z,\rho, R;t)>_{R,\rho}\,H(t)\,exp[-i\omega t]\,dt\; \mathlarger{\vert} ^2.\nonumber 
\end{eqnarray}

In order to introduce localized functions, we first decompose the total wavefunction as [13,27]
\begin{eqnarray}\label{eq:9}
  \psi(z, \rho, R;t)=& \\
   c_g(R;t)\psi_g(z, \rho;R)&+c_u(R;t)\psi_u(z, \rho;R)+\psi_{res}(z, \rho, R;t). \nonumber
\end{eqnarray}
$\psi_{g}(z, \rho;R)$ and $\psi_{u}(z, \rho;R)$ are ground and first excited electronic wavefunctions, respectively, corresponding to the $1s\sigma _g$ and $2p\sigma_u$ states. The functions $c_g(R;t)$ and $c_u(R;t)$ describe time-dependent nuclear wavepackets on the $1s\sigma _g$ and $2p\sigma_u$ states, respectively. The   $\psi_{res}(z, \rho, R;t)$ is the residual part of the $\psi(z, \rho, R;t)$  including higher excited states and electronic continuum states. We can also rewrite Eq. (9) as 
\begin{eqnarray}\label{eq:12}
  \psi(z, \rho, R;t)=a\psi_+(z, \rho;R)+b\psi_-(z, \rho;R)+\psi_{res}(z, \rho, R;t). \nonumber \\
\end{eqnarray}
with
\begin{eqnarray}\label{eq:13}
  \psi_{\pm}(z, \rho;R)=1/\sqrt{2}(\psi_g(z, \rho;R)\pm\psi_u(z, \rho;R)),
\end{eqnarray}
   
\begin{eqnarray}\label{eq:14}
  a=\sqrt{2}/2(c_g(R;t)+c_u(R;t)),
\end{eqnarray}  
\begin{eqnarray}\label{eq:15}
  b=\sqrt{2}/2(c_g(R;t)-c_u(R;t)).
\end{eqnarray} 
In these equations, $\psi_{+}(z, \rho;R)(\psi_{-}(z, \rho;R))$ is the electronic wavefunction localized on the right (left) nuclei with respect to the origin $z=0$. 
If we substitute Eq. (10) to Eq. (4) and retain the dominant terms, we arrive at 
\begin{eqnarray}\label{eq:16}
  S_{tot}\simeq S_+(\omega)+S_-(\omega)+2\text{Re}[A_+^*(\omega)\times A_-(\omega)],
\end{eqnarray}
where $S_+(\omega)=|A_+(\omega)|^2$ and $S_-(\omega)=|A_-(\omega)|^2$ and
\begin{eqnarray}\label{eq:17}
A_+(\omega)=& \\
  \int \text{2Re}< a\psi_+&(z, \rho;R)\mid a_z(t)\mid \psi_{res}(z, \rho, R;t)>H(t)e^{-i\omega t} dt,\nonumber
   \end{eqnarray}
   \begin{eqnarray}\label{eq:18}
    A_-(\omega)=& \\ \nonumber
    \int \text{2Re}< b\psi_-&(z, \rho;R)\mid a_z(t)\mid \psi_{res}(z, \rho, R;t)>H(t)e^{-i\omega t} dt. \nonumber
    \nonumber
 \end{eqnarray}
 $S_+(\omega)$ and $S_-(\omega)$ denote recombination to the $\psi_+(z, \rho;R)$ and $\psi_-(z, \rho;R)$ states, respectively, and the term 2Re$[A_+^*(\omega)\times A_-(\omega)]$ corresponds to the electronic interference term  between the localized electronic states.

\section{Results and Discussion}
 Figure 1 shows the HHG spectrum of 3D H$_2^+$ and T$_2^+$ under a 15-cycle trapezoidal laser pulse (4-7-4) at $\lambda=800$ nm and $I=3\times 10^{14}$  Wcm$^{-2}$. We have depicted the plateau region up to the cutoff region in order to focus only on recombinations from the continuum wavepacket to the $1s\sigma_g$ and $2p\sigma_u$ states. Interpretation of a HHG spectrum below the ionization potential $I_p$  ($I_p=1.1$ corresponding to harmonic order $\sim$ 20) demands  further investigation due to the transitions between the bound electronic states which  was not intended in this work.
For H$_2^+$ (Fig. 1), one can observe that odd harmonic orders are dominant. But for T$_2^+$, some non-odd harmonics and slightly red-shifted odd harmonics are seen for harmonic order 20-40.
  These differences in the spectrum pattern for the two isotopes are well understood with the consideration of effects of the falling edge of the  laser pulse, nuclear motion and ionization probability. We showed recently that even a two-cycle falling part of  a trapezoidal laser pulse leads to a significant modulation on the HHG spectrum and violation of the odd harmonic rule [13]. In order to see the contribution of the falling part of the laser pulse in HHG, the Morlet-wavelet time profile of the HHG spectra of Fig. 1  are shown for both isotopes in Fig. 2. As it is clear for H$_2^+$ in Fig. 2(a), most HHG has happened before the time $t=11$ optical cycles (o.c.) and contribution from the laser falling part ($11<t<15$ o.c.) is much lesser than the region with $t<11$ o.c. But for T$_2^+$ in Fig. 2(b), one observes considerable HHG at the falling edge of the laser pulse. The results of our simulation in this work show that the ionization probability for H$_2^+$ and T$_2^+$ at $t=11$ o.c. (at the time that laser falling part begins), is $\sim$ 56$\%$ and 6$\%$, respectively. This further population depletion of H$_2^+$ compared to T$_2^+$ is due to the charge-resonance enhanced ionization at large internuclear separations [21] which is more accessible for H$_2^+$ than T$_2^+$   because of a faster nuclear motion in the former.
Therefore, the HHG spectrum difference observed in Fig. 1 between H$_2^+$ and T$_2^+$ can arise from the induced effects of the laser falling edge. The small frequency redshift of odd harmonics in T$_2^+$ is a well-known phenomenon resulting from the non-adiabatic response of the molecular ions to the rapidly changing  field at the laser falling edge (see for more details, Refs. [6-9,12]).

 Now, we try to spatially resolve the HHG spectra in Fig. 1 to get further insight into the origin of the non-odd harmonics in T$_2^+$. In Fig. 3, we have depicted $R$-dependent harmonic profile, $S(R,\omega)$, for  H$_2^+$ (left panel) and T$_2^+$ (right panel). For H$_2^+$ (Fig. 3(a)), it is observed that HHG has been extended to the large internuclear separations up to $R \sim 8$, while for T$_2^+$ (Fig. 3(b)), HHG ends up at the lower internuclear distances around $R \sim 5.8$.  It is reasonable because  T$_2^+$ is about three times heavier than H$_2^+$ and thus it has a slower nuclear motion  under the same interaction time. Another point in Fig. 3(a) is that  odd harmonics are dominant over whole internuclear separations. But for T$_2^+$ (Fig. 3(b)), we see odd harmonic orders are dominant up to $R \sim 3.2$, but both odd and even harmonics appear for $R > 3.2$.  For H$_2^+$ that has reached to larger $R$ values than T$_2^+$, we do not observe any even harmonics because of the ionization suppression at the laser falling region. It is also obvious that larger internuclear distances ($R > 3.2$) become accessible at longer times. Above-mentioned points on Fig. 3 support  our claim that the appearance of even-order harmonics in T$_2^+$ for $R > 3.2$  is due to the effects of the falling edge of the laser pulse. 

To understand better the physics behind the non-odd harmonics in Fig. 1 for T$_2^+$, we also calculated   spatial distributions of corresponding HHG spectra as a function of the electronic coordinate $z$, $S(z,\omega)$ (left panels) and $S^+(z,\omega)$ (right panels)  which are shown in Fig. 4 for H$_2^+$ (top row) and T$_2^+$ (bottom row). $S^+(z,\omega)$ is calculated from Eq. (8) in which each $\psi(z,\rho, R;t)$ is substituted by the term $\psi(z,\rho, R;t)+\psi(-z,\rho, R;t)$. From $S^+(z,\omega)$ one can see the total HHG yield for each $|z|$ from which the spatial-dependent buildup or suppression of the harmonics can be deduced by comparing to $S(z,\omega)$. By looking at Fig. 4(a) and Fig. 4(c), it is deduced that larger $z$ values contribute into HHG for  H$_2^+$ than T$_2^+$.  It is reasonable since a larger internuclear separation becomes accessible for H$_2^+$ than T$_2^+$ due to its faster nuclear motion. 
At the origin $z=0$, one sees little probability for the harmonic emission which is also observed and demonstrated for H$_2^+$ in Ref. [29]. 
Another point in Fig. 4(a) is that HHG for two regions $z<0$ and  $z>0$ is symmetric for both odd and even harmonic orders. But in Fig. 4(c) for T$_2^+$, the HHG symmetry at positive and negative directions relative to the origin $z=0$ is broken.  $S^+(z,\omega)$ for H$_2^+$ (Fig. 4(b)) shows dominant odd harmonic orders, but for T$_2^+$ (Fig. 4(d)), besides odd harmonics, non-odd harmonics are seen which are not suppressed completely compared to the H$_2^+$.

In order to realize better how the symmetry of harmonic emission in Fig. 4 influences HHG, we have depicted $S_+(\omega)$, $S_-(\omega)$, and $S_{tot}(\omega)$ in Fig. 5 for H$_2^+$  and T$_2^+$ . It is observed for H$_2^+$ that $S_+(\omega)$ and $S_-(\omega)$ show both even and odd harmonic orders with comparable intensity (Fig. 5(a)). For $S_{tot}(\omega)$, even harmonic orders are canceled out  and odd ones are intensified. The suppression of even harmonics is due to the interference term 2Re$[A_+^*(\omega)\times A_-(\omega)]$. As harmonic emission is symmetric for H$_2^+$ in negative and positive $z$ direction (see Fig. 4(a)), HHG suppression (build-up) occurs for even (odd) harmonics based on the odd-selection rule [30]. For  T$_2^+$ in Fig. 5(b), $S_+(\omega)$ and $S_-(\omega)$ differ slightly  which is more noticeable around even harmonics. That shows that we expect to have some degrees of asymmetric emission which is compatible with asymmetry observed in Fig. 4(b). For $S_{tot}(\omega)$, one can see some even harmonics are not canceled out completely. 
 Based on above explanations, it can be concluded that harmonic peaks around even harmonic orders in Fig. 1 result from symmetry breaking in HHG on the left and right side of the simulation box around the nuclei when HHG on the falling part of the laser pulse is noticeable.  

Morales \textit{et al.} [31] also reported even-order harmonics in 1D H$_2^+$ under a 14-cycle sine-squared laser pulse at 800 nm wavelength and $I=$3 $\times 10^{14}$ Wcm$^{-2}$ intensity. They claimed that the observation of even harmonics is due to electron localization at large internuclear separation which breaks down the system's symmetry. To see whether electron localization has happened in our work, we calculated absolute asymmetry parameter $A=\text{2Re}<c_g(R;t)|c_u(R;t)>$ (see Ref. [32]) as depicted in Fig. 6, which is usually calculated to figure out the presence or absence of electron localization. As it is obvious, asymmetry parameter $A$ values not only show the same value at the end of simulations but also go to zero for both isotopes  which demonstrate that electron localization is ignorable. Therefore, in this work, the appearance of even harmonic orders cannot result from long lasting electron localization.

\section{Conclusion}
We aimed to demonstrate the induced  effects on high-order harmonic generation by the falling edge of an intense laser pulse. We investigated the interaction of molecular H$_2^+$ and T$_2^+$ isotopes under a 800 nm, 15-cycle trapezoidal laser pulse (4-7-4) of $I=$3 $\times 10^{14}$ Wcm$^{-2}$ intensity. The different time-dependent ionization suppression of these two isotopes allowed us  to observe clearly the induced effects of the 4-cycle trailing edge. In H$_2^+$, most HHG occurs before the falling edge of the laser pulse due to the high population depletion, leading to ionization suppression on the falling edge. But a substantial harmonic emission appears for T$_2^+$ on the laser trailing edge.
We found two different induced effects. The first well-known effect is non-adiabatic frequency redshift which arises due to nonadiabatic response of a medium to the rapidly changing falling edge of the laser pulse.
The second  one is  spatially asymmetric emission on the left and right side of the simulation box in the $z$ direction which induces even harmonic orders.  
In order to understand  better the asymmetric emission, we spatially resolved HHG in terms of internuclear distance and  electronic coordinate $z$ in the same direction as laser polarization one. In addition, we decompose HHG signal to different localized signal corresponding to recombination to left and right protons.  
 The spatially asymmetric emission could be considered for single- and multi-electron atomic and molecular systems as long as either one of the rising or falling parts of the laser pulse is dominant, even for  non-trapezoidal laser pulses. Also, the complicated patterns expected to appear in HHG under intense non-trapezoidal laser pulses can be understood better with the results presented in this work.

  \FloatBarrier 
\section{References}

\end{document}